\documentclass{article}
\usepackage{spconf,amsmath,graphicx,hyperref,multirow}

\title{TOWARDS HIGH-QUALITY NEURAL TTS FOR LOW-RESOURCE LANGUAGES BY LEARNING COMPACT SPEECH REPRESENTATIONS}
\name{Haohan Guo$^*$\thanks{$^*$ Work performed during the first author's internship at Xiaohongshu.}, Fenglong Xie$^\dag$, Xixin Wu$^*$, Hui Lu$^*$, Helen Meng$^*$}
\address{$^*$The Chinese University of Hong Kong, Hong Kong SAR, China \\
$^\dag$Xiaohongshu Inc., Shanghai, China \\
\ninept\href{mailto:hguo@se.cuhk.edu.hk}{\nolinkurl{{hguo, xxwu, luhui, hmmeng}@se.cuhk.edu.hk}},
  \href{mailto:fenglongxie@xiaohongshu.com}{{\nolinkurl{fenglongxie@xiaohongshu.com}}}
}

\begin{document}
%
\maketitle
\begin{abstract}

This paper aims to enhance low-resource TTS by reducing training data requirements using compact speech representations. A Multi-Stage Multi-Codebook (MSMC) VQ-GAN is trained to learn the representation, MSMCR, and decode it to waveforms. Subsequently, we train the multi-stage predictor to predict MSMCRs from the text for TTS synthesis. Moreover, we optimize the training strategy by leveraging more audio to learn MSMCRs better for low-resource languages. It selects audio from other languages using speaker similarity metric to augment the training set, and applies transfer learning to improve training quality. In MOS tests, the proposed system significantly outperforms FastSpeech and VITS in standard and low-resource scenarios, showing lower data requirements. The proposed training strategy effectively enhances MSMCRs on waveform reconstruction. It improves TTS performance further, which wins 77\% votes in the preference test for the low-resource TTS with only 15 minutes of paired data.

\end{abstract}
\section{Introduction}
\label{sec:intro}

Text-to-Speech (TTS) technologies have been widely applied to serve all people around the world in intelligent speech interactions, such as speech translation \cite{liu2021incremental,sudoh2020simultaneous}, human-machine interactions and conversations \cite{guo2021conversational}, etc. However, it becomes harder for regions using minority (even endangered) languages to achieve satisfactory TTS performance, due to the lack of training data on these languages. Hence, seeking practical approaches to address this data sparsity issue has become increasingly crucial for low-resource TTS.

Recent works on this topic mostly concentrate on leveraging more data in other fields, to compensate for the lack of target data. For example, some works \cite{45300,xu2020lrspeech,46929} aim to build a TTS dataset using crowd-sourced or automatic methods for data collection and transcribing. However, the obtained dataset may have low recording quality and low naturalness, which makes it difficult to achieve comparable performance as when standard TTS datasets are used. Therefore, some works consider using well-designed datasets in other languages to enhance TTS for low-resource languages, such as cross-lingual transfer learning \cite{azizah2020hierarchical, ChenTYL19} and multi-lingual TTS \cite{he2021multilingual}. 

Besides leveraging more data, we can also tackle this problem by reducing the training data requirement. This paper proposes learning compact speech representations to enhance low-resource TTS from this perspective. The speech waveform, as a long sequence with much redundant information, is hard to predict from the text directly without a powerful model and sufficient data \cite{Ren2021}. Hence, acoustic features with higher compactness, i.e. shorter length and fewer parameters, are usually used in TTS systems. They can be well-predicted from the text, and converted to high-fidelity waveforms via a vocoder. The compact representation effectively reduces the requirement for paired data to train acoustic models. Hence, for low-resource languages with fewer paired data, we can learn a more compact speech representation to further reduce the data requirement.

\begin{figure*}[htb]
    \centering
    \includegraphics[width=17cm]{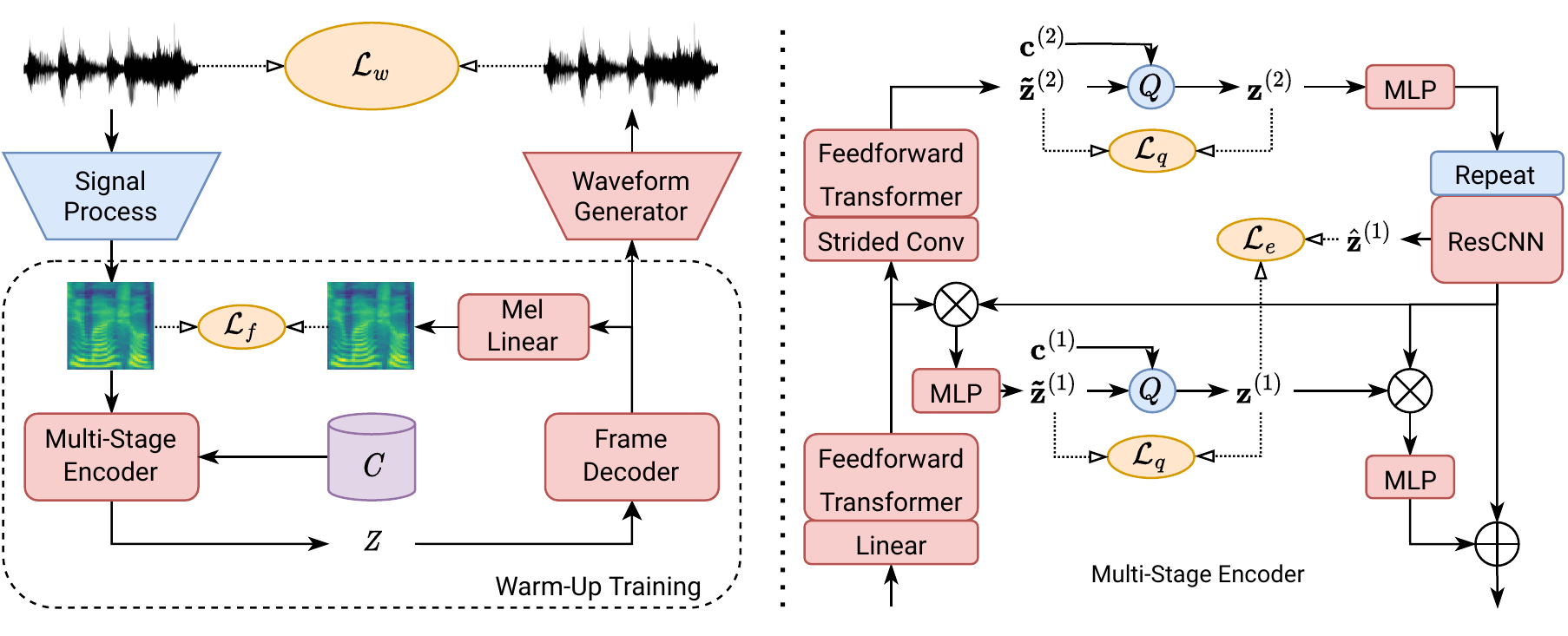}
    \caption{The model architecture of MSMC-VQ-GAN, including the framework of MSMC-VQ-GAN on the left, and the detailed structure of a 2-stage encoder in the right. $Q$, $\bigoplus$ and $\bigotimes$ denote ``quantization'', ``addition'' and ``concatenation'' operations.}
    \label{fig:msmcvqgan}
\end{figure*}

MSMC-TTS \cite{guo2022msmc} has shown great potential in this regard. It trains a Multi-Stage Multi-Codebook (MSMC) VQ-VAE to compress the waveform into the compact representation, MSMCR, i.e. a set of discrete sequences with different time resolutions. The representation can be predicted from the text by a multi-stage predictor, and converted to the waveform via a neural vocoder. In this paper, we first integrate the autoencoder and the neural vocoder into one model, MSMC-VQ-GAN, for system simplification and joint optimization. Moreover, to learn better MSMCRs for low-resource languages, we also optimize the training strategy by leveraging more high-quality audio from other languages to train MSMC-VQ-GAN. It first augments the training set by selecting utterances with high similarity to the target speaker from the low-resource language, then trains the model using transfer learning to enhance the training quality. Finally, we conduct experiments to compare the proposed system with other mainstream TTS systems under different scenarios, and evaluate the effect of the proposed training strategy on the proposed system.

\section{Method}
\label{sec:model}

\subsection{MSMC-VQ-GAN}

The optimized MSMC-TTS system comprises two models: an MSMC-VQ-GAN based autoencoder, and a multi-stage predictor as the acoustic model. Inspired by VQ-GAN \cite{esser2021taming}, MSMC-VQ-GAN integrates MSMC-VQ-VAE \cite{guo2022msmc} and HiFi-GAN vocoder \cite{Kong2020} together for both representation learning and waveform generation. Fig. \ref{fig:msmcvqgan} shows the model architecture of MSMC-VQ-GAN. The input speech waveform is first converted to the Mel spectrogram by signal processing. The multi-stage encoder encodes the spectrogram stage-wise into multiple sequences with different time resolutions. We then uses multi-head vector-quantization, i.e. product quantization \cite{jegou2010product}, to quantize these hidden sequences respectively in a reversed order. For example, for the 2-stage encoder, the Mel spectrogram is first encoded by a linear layer and a feedforward transformer block to the first hidden sequence. Then, the second hidden sequence with a lower time resolution is generated by a strided convolutional layer for down-sampling and the second transformer block for further processing. The output sequence is quantized by the codebook $\mathbf{c}^{(2)}$ to obtain $\mathbf{z}^{(2)}$, which is then processed by the MLP layer and a residual CNN module to help obtain $\mathbf{z}^{(1)}$. These two quantized sequences forms MSMCR $\mathit{Z} = \{\mathbf{z}^{(1)}, \mathbf{z}^{(2)}\}$. Finally, the residual output of the encoder is fed to the transformer block based frame decoder and the HiFi-GAN based waveform generator to reconstruct the waveform.

Adversarial training with a UnivNet discriminator \cite{Jang2021} is applied to train the model. The loss function is written as follows:
\begin{equation}
    \mathcal{L} = \lambda_w * \mathcal{L}_{w} + \lambda_f * \mathcal{L}_{f} + \lambda_q * \mathcal{L}_{q} + \lambda_e * \mathcal{L}_{e}
\end{equation}
$\mathcal{L}_{w}$ is the waveform-level loss between the ground-truth waveform and the reconstructed one, and is implemented with the loss function of HiFi-GAN. The frame-level loss $\mathcal{L}_{f}$ calculates the MSE between two Mel spectrograms. The VQ loss $\mathcal{L}_{q}$ back-propagates gradient to the encoder by minimizing the MSE loss between $\mathbf{\tilde{z}}^{(i)}$ and its quantized output $\mathbf{z}^{(i)}$. The last term $\mathcal{L}_{e}$ is applied to stabilize the training process by minimizing the MSE between the ground-truth latent sequence $\mathbf{z}^{(i)}$ and the predicted quantized sequence $\mathbf{\hat{z}}^{(i)}$ from the higher stage. $\lambda_w, \lambda_f, \lambda_q, \lambda_e$ are weight coefficients to balance these terms. In training, a batch of utterances is fed to the model to calculate $\mathcal{L}_{e}, \mathcal{L}_{q}, \mathcal{L}_{f}$, but only a few segments sampled from the output of the frame decoder are converted to waveforms to calculate $\mathcal{L}_{w}$ in adversarial training. It effectively reduces the memory cost, and speeds up the training process, widely used in adversarial end-to-end TTS \cite{kim2021conditional}. Moreover, warm-up training without $\mathcal{L}_{w}$ is applied in the first thousands of iterations to stabilize the training process. VQ codebooks are updated using the exponential moving average based method \cite{razavi2019generating} in this work.

After the training of MSMC-VQ-GAN is completed, MSMCRs are extracted from the training set to train the acoustic model, which is a multi-stage predictor proposed in MSMC-TTS \cite{guo2022msmc}. It maps the text to multiple sequences with different time resolutions, and is trained with the loss function combining MSE and ``triplet loss'' to find expected codewords in the continuous space. 

\subsection{Training Strategy}
\label{ssec:train}

The compact representation reduces the training data requirement of the acoustic model. Hence, we can use fewer paired data with text and audio for training. However, it is also critical to ensure high-quality waveform reconstruction from the representation, i.e. high feature completeness. Hence, to learn better MSMCRs for low-resource languages, we refer to mainstream data-driven methods, and propose a training strategy leveraging more audio from other languages to train MSMC-VQ-GAN, which is composed of two parts: data selection and transfer learning.

First, we propose augmenting the training set by selecting utterances from other languages. Datasets from different languages share common knowledge in different aspects, such as cross-lingual phonetic information \cite{international1999handbook} and the speaker-dependent attributes, which are helpful for low-resource TTS \cite{ChenTYL19}. We can also leverage these data to train MSMC-VQ-GAN. However, it is critical to seek a practical ranking criterion to select valuable utterances from the multi-lingual candidate set. Otherwise, utterances with significant differences from the target-lingual dataset may contaminate the model instead, causing covariate shift \cite{quinonero2008dataset}, degrading MSMCRs in representing target-lingual speech. The differences are usually reflected in speaker-dependent attributes, e.g. timbre and speaking style. Hence, we propose sorting these utterances in terms of speaker similarity. A well-trained ECAPA-TDNN \cite{desplanques2020ecapa} based speaker verification model extracts the speaker embedding for each candidate utterance. Then, we rank these utterances according to their cosine similarity with the target speaker. Finally, top-$k$ utterances are selected to couple with the target dataset to form the multi-lingual training set. 

Although unselected candidate utterances have significant differences in speaker similarity, they still can provide valuable knowledge in other aspects, e.g. phonetics, prosody, etc. Hence, we apply transfer learning to further leverage these data. We first use the whole candidate set to train MSMC-VQ-GAN. Then, for the low-resource language, the pre-trained MSMC-VQ-GAN will be fine-tuned with the training set augmented with the proposed data selection method. In this way, we can transfer the knowledge learned from the large-scale multi-lingual dataset to low-resource languages to further enhance the model. Furthermore, this method also reduces the training cost for low-resource TTS. The pre-trained MSMC-VQ-GAN can be used for arbitrary languages by fine-tuning with much fewer iterations than from-scratch training. In conclusion, based on this training strategy, we can learn expected MSMCRs faster and better for low-resource languages.

\section{Experiments}
\label{sec:exp}

\subsection{Experimental Setup}

In this paper, we conduct experiments on CSMSC, a standard Chinese TTS dataset with 10 hours of paired data, and set up low-resource scenarios by randomly selecting 1,000 (1 hour) and 250 (15 minutes) utterances from this dataset. The multi-speaker English dataset, LibriTTS \cite{zen2019libritts}, is employed as the candidate set. The test set is composed of 200 utterances from CSMSC but out of all training sets. All audios are down-sampled to the 24k sample rate, and converted to 80-dim log-scale Mel spectrograms with a frameshift of 12.5 ms. Chinese characters are converted to the phoneme sequences as the input of the acoustic model. MSMC-VQ-GAN is implemented with 256-dim feedforward transformer blocks \cite{Ren2019} and HiFi-GAN-V1 as the waveform generator. It extracts the 2-stage MSMCR with the down-sample rates 1 and 4, where each stage uses the 4-head vector-quantization with 64 codewords per head. All MSMC-VQ-GAN models are trained for 800,000 iterations, including 50,000 warm-up iterations, with a batch size of 16 utterances, where the segment with the length of 0.5 seconds is randomly sampled from each utterance for adversarial training. The acoustic model is implemented with the same configurations in \cite{guo2022msmc}, and trained for 200,000 iterations with a batch size of 64.\footnote{The detailed implementation is available in the released code of MSMC-TTS at \url{https://github.com/hhguo/MSMC-TTS}. Audio samples are available at \url{https://hhguo.github.io/DemoMSMCLRTTS}.}

\begin{table}[htp]
\centering
\caption{MOS test: TTS system comparison}
\label{tab:mos_sys}
\begin{tabular}{c|cc}
\hline
\textbf{\# Utterances} & \textbf{Model} & \textbf{MOS ($\pm$ 95\% CI)} \\ \hline
- & Recording & 4.48 $\pm$ 0.12 \\ \hline
\multirow{3}{*}{9,800} & FastSpeech & 3.52 $\pm$ 0.15 \\
 & VITS & 3.98 $\pm$ 0.13 \\
 & MSMC-TTS & 4.21 $\pm$ 0.12 \\ \hline
\multirow{3}{*}{1,000} & FastSpeech & 1.74 $\pm$ 0.11 \\
 & VITS & 1.51 $\pm$ 0.09 \\
 & MSMC-TTS & 3.55 $\pm$ 0.13 \\ \hline
\end{tabular}%
\end{table}

\subsection{System Comparison}

We first evaluate the proposed TTS system in standard and low-resource scenarios by comparing it with other mainstream systems, FastSpeech \cite{Ren2019} with a HiFi-GAN vocoder, and VITS \cite{kim2021conditional}. We conduct the MOS test by employing ten native speakers to score each audio in 20 test cases from 1 to 5 with an incremental of 0.5. Table \ref{tab:mos_sys} shows the test result.

First, in the standard scenario with 9,800 training pairs, the end-to-end approach, VITS, using much more training cost, i.e. 1,200k iterations on 4 GPUs, achieves better performance than FastSpeech. However, MSMC-TTS trained with only 1 GPU shows the best performance with an MOS of 4.21, outperforming FastSpeech and VITS significantly. This validates the effectiveness of the proposed TTS system. Furthermore, this improvement is more prominent in the low-resource scenario with only 1,000 training pairs. While FastSpeech and VITS fail to build reliable TTS systems without sufficient data, MSMC-TTS still generates high-quality speech, even outperforming FastSpeech using the standard dataset. It strongly verifies the lower data requirement of MSMC-TTS, and its superior performance in the low-resource scenario.

\subsection{Evaluation of Proposed Training Strategy}

\subsubsection{Evaluation of Analysis-Synthesis}

\begin{figure}[htp]
    \centering
    \includegraphics[width=8.5cm]{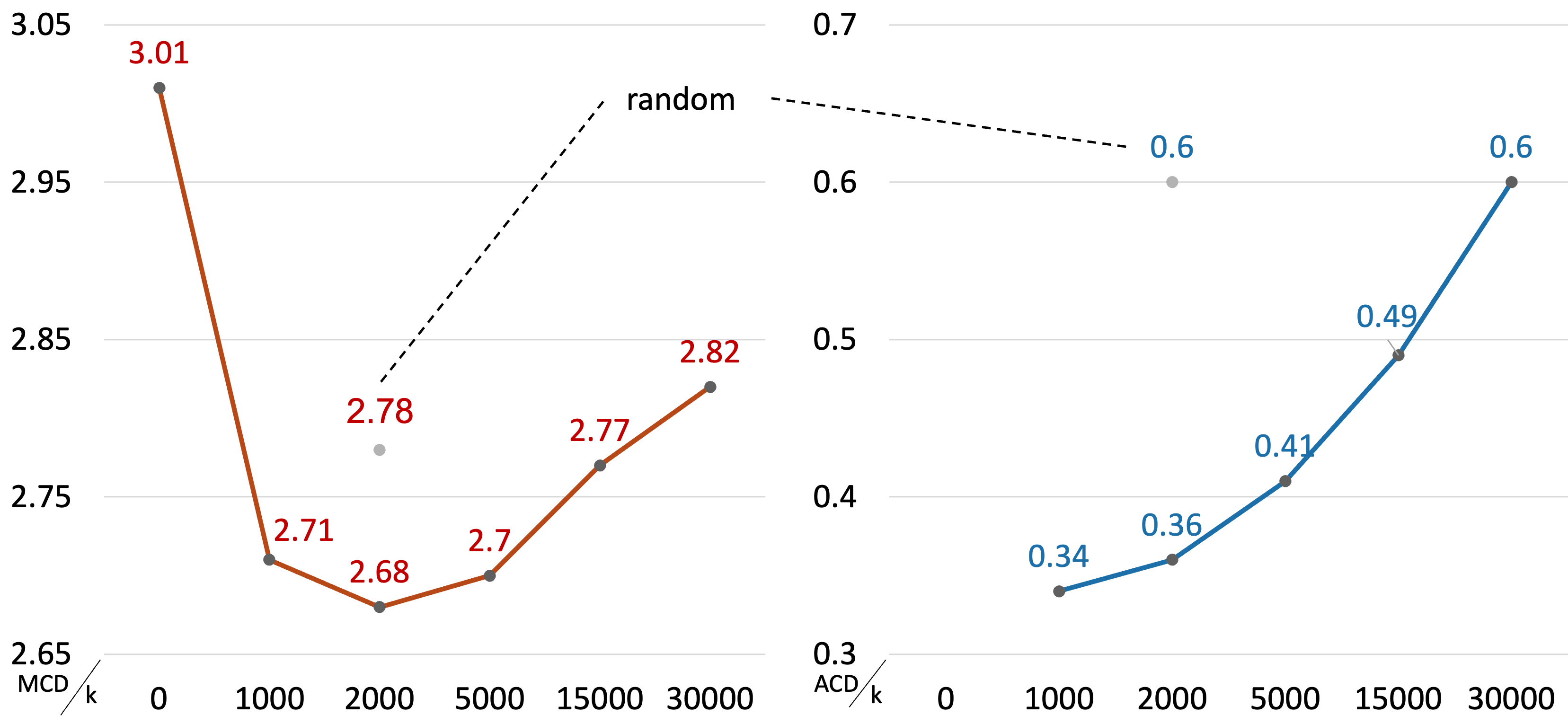}
    \caption{MCD (left) of models trained with top-$k$ utterances and Averaged Cosine Distance (right) of top-$k$ utterances}
    \label{fig:chart}
\end{figure}

We first evaluate the training strategy using analysis by synthesis, i.e. reconstructing the waveform from the MSMCR. For data selection, we compare models trained from scratch using 1,000 utterances from CSMSC and top-$k$ utterances sorted with regard to speaker similarity from LibriTTS. As shown in Fig. \ref{fig:chart}, the increased Averaged Cosine Distance (ACD) shows the lower speaker similarity as more utterances are selected. The red curve illustrates the change of reconstruction quality (Mel Cepstral Distortion, MCD) of models on the test set as $k$ increases. First, the model trained without data selection, i.e. $k=0$, obtains the highest MCD of 3.03 dB. Given the top-1000 utterances, MCD significantly decreases to 2.71 dB, and further improves to 2.68 dB with 1,000 more utterances. It validates the effectiveness of employing more audio from other languages. However, with $k$ further increasing, the model degrades instead, since low-similarity utterances are increasingly introduced, and contaminates the training. Hence, selecting utterances with high similarity to the target speaker is essential. The model using 2,000 randomly selected utterances further verifies this claim. These utterances have a higher ACD of $0.6$, giving less improvement to the model by an MCD of 2.78 dB, worse than the model using top-2000 utterances.

We then evaluate the effect of transfer learning, which first pre-trains MSMC-VQ-GAN with the whole LibriTTS, then fine-tunes it on created datasets for 200,000 iterations. As shown in Table \ref{tab:obj_transfer}, transfer learning can significantly improve the reconstruction quality compared with models trained from scratch using only 1,000 or 250 utterances from CSMSC. Given the training set augmented with 2,000 utterances selected via the proposed method, we obtain higher-quality models in both situations, especially for the extremely low-resource scenario with only 250 utterances, obtaining more improvement with the MCD of 2.70 dB and F0-RMSE of 1.89 Hz. To summarize, the proposed training strategy shows its effectiveness in enhancing MSMC-VQ-GAN, and is more prominent with lower resources in the target language.

\begin{table}[htp]
\centering
\caption{Objective metrics: reconstruction quality of different MSMC-VQ-GANs (``TL'' and ``DS'' denote whether to use transfer learning and data selection, respectively)}
\label{tab:obj_transfer}
\begin{tabular}{ccc|ccc}
\hline
\textbf{\# Utts} & \textbf{TL} & \textbf{DS} & \textbf{\begin{tabular}[c]{@{}c@{}}MCD\\ (dB)\end{tabular}} & \textbf{\begin{tabular}[c]{@{}c@{}}F0-RMSE\\ (Hz)\end{tabular}} & \textbf{\begin{tabular}[c]{@{}c@{}}F0-VUV\\ (\%)\end{tabular}} \\ \hline

\multirow{3}{*}{1,000} & No & No & 3.01 & 2.45 & 3.75 \\
  & Yes & No & 2.61 & 1.84 & 2.83 \\
 & Yes & Yes & 2.59 & 1.67 & 2.70 \\ \hline
\multirow{3}{*}{250} & No & No & 4.49 & 3.31 & 7.29 \\
 & Yes & No & 2.97 & 2.48 & 3.75 \\
 & Yes & Yes & 2.70 & 1.89 & 2.92 \\ \hline
\end{tabular}
\end{table}

\subsubsection{Evaluation of TTS Synthesis}

Finally, we compare TTS systems trained with or without the proposed training strategy via preference tests. As shown in Table \ref{tab:abtest}, under the situation with 1,000 utterances from CSMSC, the training strategy improves TTS quality by the preference of 34.6\%, showing higher naturalness and fidelity. In the extremely low-resource situation, MSMC-TTS trained with only 250 utterances has severe problems in both intelligibility and audio fidelity. However, the training strategy significantly enhances the TTS system, which produces intelligible and natural speech, obtaining a much higher preference of 77.1\%. Hence, by leveraging more audio from other languages to enhance the MSMC-VQ-GAN, we can obtain better MSMCRs, and effectively enhance low-resource TTS without using more paired data.

\begin{table}[htb]
\centering
\caption{ABTest: MSMC-TTS trained with or without the proposed training strategy}
\label{tab:abtest}
\begin{tabular}{c|ccc}
\hline
\textbf{\# Utterances} & \textbf{w/o} & \textbf{on par} & \textbf{w/} \\ \hline
1,000 & 28.3 & 37.1 & 34.6 \\
250 & 2.9 & 20.0 & 77.1 \\ \hline
\end{tabular}
\end{table}

\section{Conclusions}
\label{sec:con}

This paper purposes to enhance low-resource TTS with compact representations. Based on MSMC-TTS, we propose MSMC-VQ-GAN to learn MSMCR to replace acoustic features for TTS synthesis, and reconstruct the waveform from it. To learn better MSMCRs for low-resource languages, we optimize the training strategy using more audio from other langauges. It augments the training set by selecting utterances w.r.t. speaker similarity, and applies transfer learning in training. In experiments, the proposed system outperforms baseline systems significantly in standard and low-resource scenarios. Furthermore, the training strategy shows its effectiveness in learning MSMCRs, enhancing low-resource TTS significantly, which generates intelligible and natural speech with only 1 hour or 15 minutes of paired data.

\vfill\pagebreak

\bibliographystyle{IEEEbib}
\bibliography{refs}

\end{document}